# (111) Facet-engineered SnO$_2$ as Electron Transport Layer for Efficient and Stable Triple-Cation Perovskite Solar Cells

*Keshav Kumar Sharma, Rohit, Sochannao Machinao, and Ramesh Karuppannan*

**Department of physics, Indian Institute of Science, Bangalore 560012, Karnataka, India**

Corresponding Author: <u>kramesh@iisc.ac.in</u>

**ABSTRACT**: We report the (111) facet-engineered cubic phase SnO$_2$ (C-SnO$_2$) as a novel electron transport layer (ETL) for triple-cation mixed-halide Cs$_{0.05}$(FA$_{0.83}$MA$_{0.17}$)$_{0.95}$Pb(I$_{0.83}$Br$_{0.17}$)$_3$ perovskite solar cells (PSCs). The C-SnO$_2$ layer was prepared via a normal sol-gel process followed by the spin-coating technique. The (111) facet C-SnO$_2$ layer provides a larger surface contact area with an adjacent perovskite layer, enhancing charge transfer dynamics at the interface. In addition, the well-matched overlapping band structures improve the charge extraction efficiency between the two layers. Using (111) facet C-SnO$_2$ as ETLs, we obtain PSCs with a higher power conversion efficiency of 20.34% (0.09 cm$^2$) compared to those employing tetragonal phase SnO$_2$ ETL. The PSCs with C-SnO$_2$ ETL retain over 81% of their initial efficiency even after 480 h. This work concludes with a brief discussion on recombination and charge transport mechanisms, providing ways to optimize C-SnO$_2$ ETL leads to better performance and stability of the PSCs.

**INTRODUCTION**
State-of-the-art perovskite solar cells (PSCs) have emerged as a promising technology for renewable energy generation, owning to their outstanding power conversion efficiency (PCE), wide spectral absorption, large carrier lifetime, and facile band gap engineering[1–3]. PSCs have shown remarkable progress in PCE, from 3.8% in 2009 to over 26.08% (certified 25.73%) in less than 15 years[4,5]. However, PSCs based on single or dual cation perovskites, such as methylammonium lead perovskite (MAPbI$_3$) or formamidinium lead iodide (FAPbI$_3$), face several challenges that hinder their commercialization, such as thermal instability, moisture sensitivity, phase separation, and narrow optimal processing conditions[6–10]. To address these issues, triple cation mixed halide (Cs$_{0.05}$(MA$_{0.17}$FA$_{0.83}$)$_{0.95}$Pb(I$_{0.83}$Br$_{0.17}$)$_3$, abbreviated as CsFAMA) perovskites have been introduced[11–13]. The performance of perovskite solar cells (PSCs) is highly dependent on the quality of the perovskite absorbing layer. The crystallization process and crystallinity of the absorber layer are influenced by the adjacent charge transporting layer. This interface layer controls charge transfer kinetics and affects defect-induced charge trap states.

Various wide band gap semiconducting materials such as TiO$_2$, ZnO, and SnO$_2$ have been used as ETL to improve the optoelectronic properties of the PSCs. These materials provide reduced shunt resistance and larger surface contact area between the transparent electrode and perovskite interface, improving the charge transfer dynamics at the corresponding interface[14–17]. SnO$_2$ ETL has been mostly used in PSCs due to its high transparency, stability, low temperature processibility, and excellent performance in terms of efficiency and stability compared to TiO$_2$ and ZnO[18,19,20]. Zakiria et al. fabricated the triple cation PSCs of 17.10% PCE using RF magnetron sputtered SnO$_2$ ETL[21]. Lee et al. reported the dopant-free, amorphous-crystalline heterophase SnO$_2$ ETL >20% PCE of PSCs, which exhibits improved surface morphology, considerably fewer oxygen defects, and better energy band alignment with triple-cation perovskite without scarifying the optical transmittance[22]. Zeng et al. introduced double-halide composition-engineered SnO$_2$ ETL for triple-cation PSCs with 20.3% efficiency[23]. Lou et al. reported π-conjugated small molecules modified SnO$_2$ ETL for triple-cation PSCs with over 23% efficiency. Despite these advantages, key challenging issues such as poor crystallinity, lower charge extraction rate and surface defects-induced small diffusion length, still need to be addressed. In this work, we introduce dopant-free (111) facet C-SnO$_2$ layer as ETL for triple cation PSCs to addressed above issues. The C-SnO$_2$ is a high-pressure phase in bulk form and is not stable at ambient conditions. In our earlier work, we have shown that the thin films of (111)-oriented C-SnO$_2$ can be prepared at ambient conditions with an excellent stability[24].

In this study, we report the synthesis of (111) facet C-SnO$_2$ thin films and the performance of PSCs using C-SnO$_2$ as an alternative ETL material. An important advantage of C-SnO$_2$ is its ability to crystallize at lower temperature (150 °C). The proposed C-SnO$_2$ ETL can provide more surface contact with the CsFAMA overlayer due to surface polarity and therefore enhance the charge extraction efficiency at the SnO$_2$/CsFAMA interface. Furthermore, the C-SnO$_2$ ETL exhibits better optical band alignment with CsFAMA compared to T-SnO$_2$, which also helps to improve the charge extraction efficiency. With these advantages, the PSC with C-SnO$_2$ ETL shows a maximum PCE of 20.34% for device area of 0.09 cm$^2$ whereas the PSC with T-SnO$_2$ exhibits a PCE of 19.64%. Moreover, the PSCs with C-SnO$_2$ maintain 81% of initial PCE after 480 h of light illumination.

**RESULTS AND DISCUSSION**

Cubic phase SnO$_2$ was synthesized by a normal sol-gel method followed by the spin coating technique, as detailed in the experimental methods section. First, Tin(IV) chloride pentahydrate (SnCl$_4$.5H$_2$O) powder was dissolved in pure deionized (DI) water. The SnO$_2$ precursor was then stirred at the boiling point of water. A transparent colloidal precursor was formed. We used phase analysis light scattering (PALS) to measure the particle size distribution of the SnO$_2$ precursor solution (Figure 1a). The particle size in fresh SnO$_2$ precursor exhibited a typical Gaussian distribution with the peak cluster observed at ~51.12 nm, while a greater peak position (~182.10 nm) was observed for the same precursor after 24 hours of aging. Solution aging inhibits the aggregation of colloids inside, thereby enhancing the stability of the SnO$_2$ precursor and improving the uniformity of the prepared SnO$_2$ ETL.

The $^{119}$Sn Mössbauer spectrum of the powder sample extracted from the SnO$_2$ precursor was recorded at room temperature, as shown in Figure 1b. The $^{119}$Sn spectrum exhibited a single line with isomer shift 0.06 mm/s, indicating the presence of only the Sn$^{4+}$ state. The FTIR spectrum of the SnO$_2$ precursor is shown in Figure 1c. The bands at 490 and 924 cm$^{-1}$ are assigned to the Sn–O stretching modes[25,26]. The band at 606 cm$^{-1}$ is related to the O–Sn–O bending modes of vibration[27,28]. The bands at 1102 and 1612 cm$^{-1}$ are ascribed to the Sn–OH stretching and H–O–H bending modes of vibration, respectively[25–28]. The broad bands at 2448 and 3358 cm$^{-1}$ are assigned to the C–H stretching mode and the O–H stretching vibration of water molecules[26,27]. Further, X-ray diffraction of powder extracted from SnO$_2$ precursor was collected (Figure 1d), confirming the pure tin(IV) oxide-hydroxide (SnO(OH)$_2$)[29,30].

The HR-TEM image of SnO$_2$ nanoparticles (NPs), depicted in Figure 1e, shows that the SnO$_2$ NPs exhibit highly crystalline, closely packed grains with an average size ranging from 10 to 20 nm. The inset in Figure 2e shows the HR-TEM image of the selected area marked by the square in Figure 2e, showing clearly visible and parallel lattice fringes, which confirm the single oriented crystalline nature of the SnO$_2$ NPs. The calculated interplanar spacing of 0.283 nm corresponds to the d-spacing of the (111) planes of the pyrite cubic SnO$_2$. The selected-area electron diffraction (SAED) pattern of the SnO$_2$ NPs, obtained from HR-TEM analysis and shown in Figure 1f, indicates that the $d_{hkl}$ values of various {hkl} planes in the SAED pattern match those of the cubic phase (space group *Pa-3*)[24].

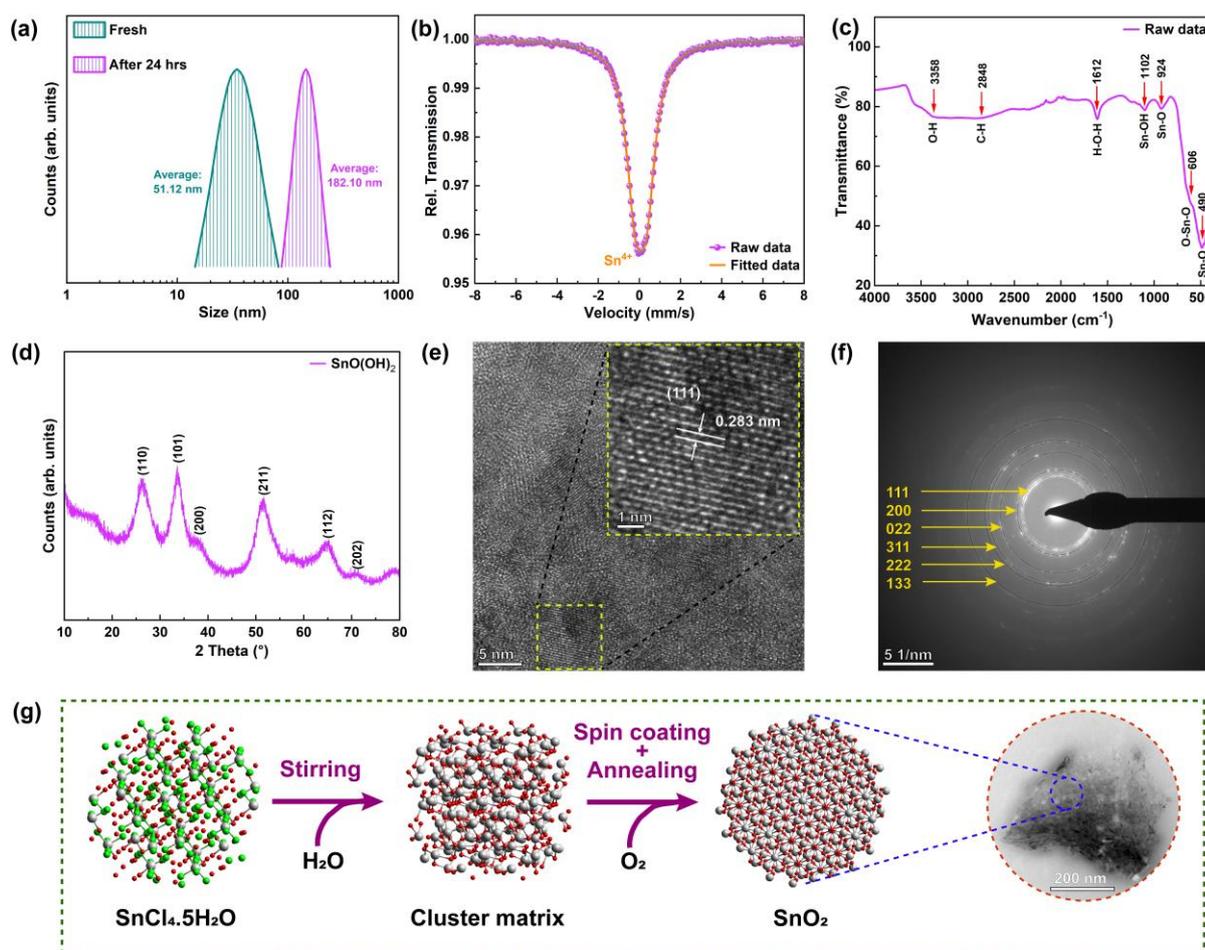

**Figure 1:** (a) Particle size distribution of SnO$_2$ precursor solution, (b) $^{119}$Sn Mössbauer spectrum, (c) FTIR spectrum, (d) XRD pattern of SnO$_2$ precursor, (e) HR-TEM image, (f) SAED pattern of cubic phase SnO$_2$, and (g) Schematic illustration of the formation mechanism of C-SnO$_2$ thin film.

The formation mechanism of the C-SnO$_2$ thin film is shown in Figure 1g. At first, Tin(IV) chloride pentahydrate and water produce Sn$^{4+}$ and OH$^-$, respectively, through ionization in the solution. The Sn$^{4+}$ and OH$^-$ ions react to form Sn(OH)$_4$. The dehydration of Sn(OH)$_4$ is quite complicated and creates cluster matrix in the solution[31]. Later, upon spin coating this solution and subsequently annealing it, C-SnO$_2$ nanoparticles are formed.

The X-ray diffraction (XRD) patterns of the T-SnO$_2$ and C-SnO$_2$ films are shown in Figure 2a. For the tetragonal phase, the XRD pattern typically exhibits prominent peaks at (110), (101), and (211), which are indicative of the rutile-type crystalline structure with the space group *P4$_2$/mnm*[32,33]. Conversely, the XRD pattern for the cubic phase shows characteristic peaks at (111) and (222), corresponding to the pyrite-type crystalline structure with the space group *Pa-3*[24]. Subsequently, we tested the conductivity of T-SnO$_2$ and C-SnO$_2$ samples using the dark *I–V* curve of a simple device, as depicted in Figure S1. The C-SnO$_2$ film exhibited an increase in electrical conductivity compared to the T-SnO$_2$ film. This higher conductivity may be due to the fewer defects in the C-SnO$_2$ film.

The X-ray photoelectron spectroscopy (XPS) analysis of SnO$_2$ was conducted to examine the surface compositions and their respective valence states. Sn *3d* and O *1s* XPS spectra of both tetragonal and cubic phase SnO$_2$ NPs thin films are shown in Fig 2b,c. The XPS spectra of T-SnO$_2$ reveal the Sn 3d$_{5/2}$ and Sn 3d$_{3/2}$ peaks at binding energies of 487.01 eV and 495.44 eV, respectively. In contrast, the corresponding peaks for C-SnO$_2$ are observed at 487.05 eV and 495.48 eV. These binding energies are indicative of the Sn$^{4+}$ oxidation state. No additional peaks corresponding to other oxidation states, such as $Sn^{2+}$ or $Sn^0$, were detected[34,35]. The O $1s$ spectrum of T-SnO$_2$ was deconvoluted into two Gaussian component peaks, designated as O$_I$ and O$_{II}$, with binding energies at 530.76 and 532.03 eV, respectively. The O$_I$ peak is typically associated with the octahedral coordination of [SnO$_6$] lattice oxygen states in T-SnO$_2$, which includes superoxide anions ($O^{2-}$)[36,37]. The O$_{II}$ peak, centered at 532.03 eV, can be attributed to the photoejection of $1s$ electrons from oxygen atoms within chemisorbed (strongly bound) water molecules on the surface. Additionally, surface oxygen atoms passivated with adsorbed hydrogen may also contribute to the 532.03 eV binding energy photoemission[38,39]. The O $1s$ spectrum of C-SnO$_2$ was deconvoluted into three Gaussian component peaks, designated as O$_I$, O$_{II}$, and O$_{III}$, with binding energies at 530.78, 532.05 and 533.20 eV, respectively. The additional O$_{III}$ peak of O $1s$ spectrum of C-SnO$_2$ is associated with more loosely bound water molecules, which contain core electrons corresponding to a binding energy of 533.20 eV[38].

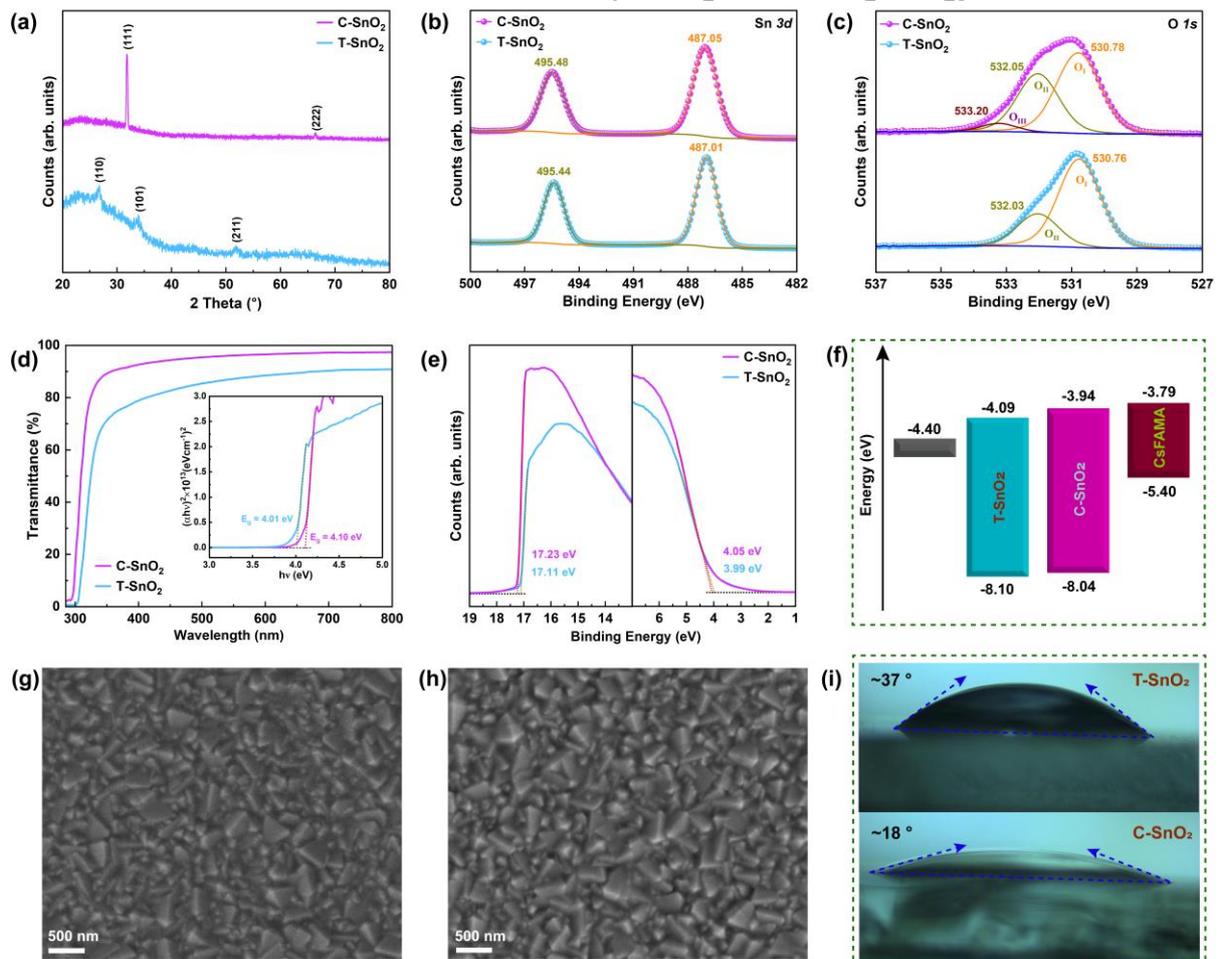

**Figure 2:** (a) The XRD patterns, (b) Sn 3d XPS spectra, (c) O 1s XPS spectra, (d) transmittance spectra with Tauc plots, (e) UPS spectra of T-SnO$_2$ and C-SnO$_2$ films, (f) schematic diagram of energy level structure of FTO/SnO$_2$/CsFAMA, (g, h) SEM images, and (i) water contact angles of the T-SnO$_2$ and C-SnO$_2$ films.

The energy band structures of T-SnO$_2$ and C-SnO$_2$ films were meticulously analyzed using ultraviolet-visible (UV-vis) spectroscopy and ultraviolet photoelectron spectroscopy (UPS). The transmittance spectra, depicted in Figure 2d, were employed to estimate the optical band gaps of these films. C-SnO$_2$ film exhibits high transmittance. By applying the Tauc plot method (Figure 2d), the optical band gaps of T-SnO$_2$ and C-SnO$_2$ were determined to be 4.01 eV and 4.10 eV, respectively. This precise measurement highlights the slight variation in the electronic properties of the two polymorphs, which can be attributed to their distinct crystal structures.

The UPS spectra of T-SnO$_2$ and C-SnO$_2$ films are illustrated in Figure 2e. Detailed analysis of these spectra indicates that the conduction band minima (CBM) of the C-SnO$_2$ film is situated at a slightly higher energy level compared to that of the T-SnO$_2$ film (Figure 2f). Additionally, the UV-vis absorption and UPS spectra of the CsFAMA film are depicted in Figure S2. Based on these observations, the elevated conduction band in the C-SnO$_2$ film proves advantageous as it mitigates the voltage loss at the interface between the C-SnO$_2$ film and the perovskite layer within the device. This favorable alignment enhances electron injection efficiency, thereby significantly improving the overall performance of the devices.

The surface morphology of the SnO$_2$ ETL plays a significant role in influencing the nucleation and growth of the perovskite crystals. The SEM images of T-SnO$_2$ and C-SnO$_2$ films are shown in Figure 2g,h. The C-SnO$_2$ film shows the formation of large, dense nanoparticles, and a uniform surface. In addition, we measured the root-mean-square (RMS) roughness of the T-SnO$_2$ and C-SnO$_2$ films by using atomic force microscopy (AFM). As shown in Figure S2, the RMS roughness of the C-SnO$_2$ film was measured to be 7.83 nm, which is lower than that of the T-SnO$_2$ film (10.62 nm). The water droplet angle test, as depicted in Figure 2i, showed that the contact angle of the T-SnO$_2$ film is 37°, while it reduces to 18° for C-SnO$_2$ film. This reduction in contact angle may be due to the higher surface polarity of the C-SnO$_2$ film[24]. The smoother and more hydrophobic surface of the C-SnO$_2$ film is more conducive to reducing nucleation sites, promoting crystal growth and the formation of CsFAMA perovskite layer with larger grains on it[40].

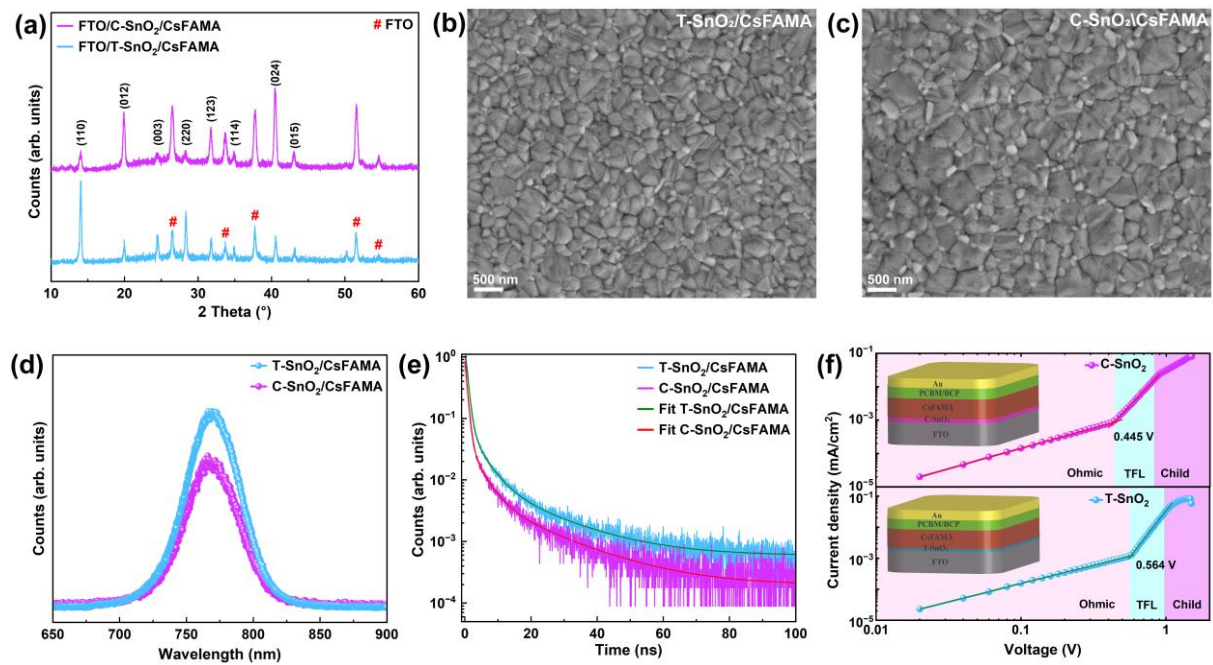

**Figure 3:** (a) XRD patterns of the CsFAMA films on the T-SnO$_2$ and C-SnO$_2$ films, (b,c) SEM images of CsFAMA films on the T-SnO$_2$ and C-SnO$_2$ ETLs, (d,e) PL and TRPL spectra of the CsFAMA films on the T-SnO$_2$ and C-SnO$_2$ ETLs, and (f) $J$—$V$ characteristics of the electron-only devices.

The XRD patterns of CsFAMA films showed that the intensity of diffraction peaks of CsFAMA film on the C-SnO$_2$ layer is different to those of the CsFAMA on the T-SnO$_2$ layer (Figure 3a). This variation in the peak intensities may be due to the smother and more hydrophobic surface of the C-SnO$_2$ layer. The scanning electron microscopy (SEM) images of CsFAMA films deposited on the T-SnO$_2$ and C-SnO$_2$ layers, are shown in Figure 3b and 3c, respectively. The morphology results show the formation of larger nanoparticles and uniform CsFAMA film on the C-SnO$_2$ layer.

To investigate the effect of C-SnO$_2$ on the charge dynamics within the CsFAMA perovskite film, we initially collected the photoluminescence (PL) and time-resolved PL (TRPL) spectra of CsFAMA films deposited on both T-SnO$_2$ and C-SnO$_2$ ETLs. The PL intensity significantly decreased for C-SnO$_2$/CsFAMA, as shown in Figure 3d. The TRPL spectra of CsFAMA, shown in Figure 3e, reveals that the average PL lifetimes were 7.38 ns and 3.21 ns for T-SnO$_2$ and C-SnO$_2$ ETLs, respectively. The PL and TRPL spectra of pristine CsFAMA film are shown in Figure S4 and S5. Furthermore, the density of trap states were calculated to be $1.80 \times 10^{16} cm^{-3}$ (T-SnO$_2$) and $1.42 \times 10^{16} cm^{-3}$ (C-SnO$_2$) using space-charge-limited current (SCLC) method[40,41] on the electron-only devices (Figure 3f). The observed reduction in PL intensity, along with the reduced PL lifetime and trap state density for the C-SnO$_2$ ETL collectively contributed to the improved charge extraction and reduced charge recombination rates in the PSCs, confirming that the C-SnO$_2$ layer effectively modulates the crystallinity and charge recombination dynamics of the CsFAMA perovskite film[40,41].

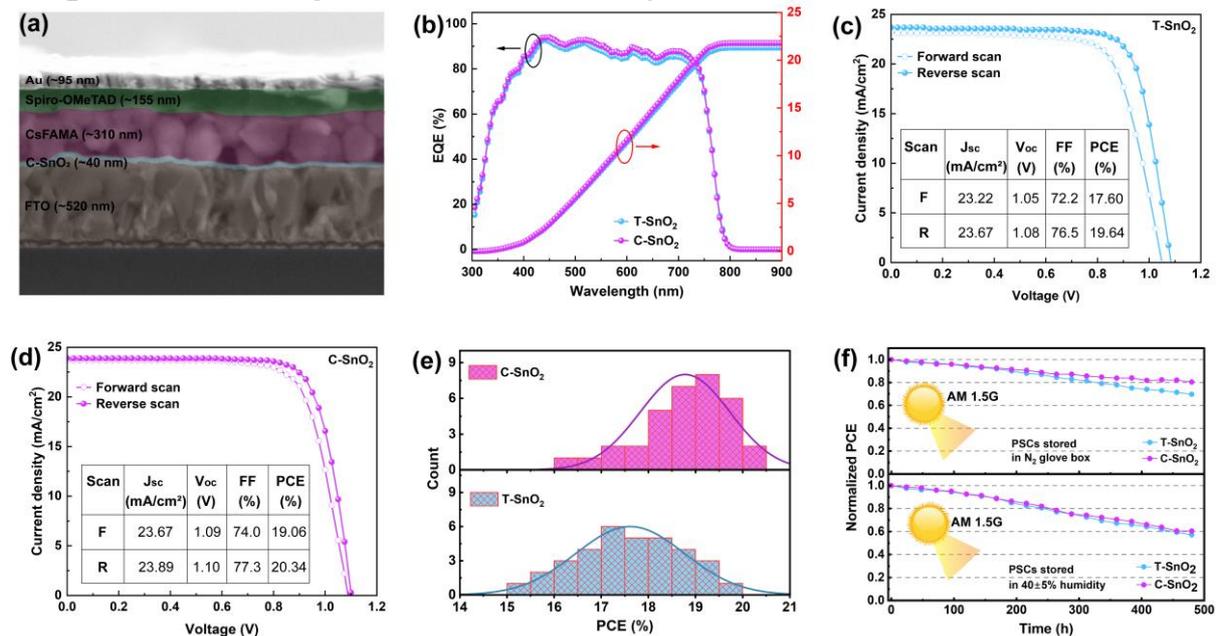

**Figure 4:** (a) Cross-section SEM image of the PSC, (b) EQE and integrated curves, (c) forward and reverse scan J-V curves of PSC with T-SnO$_2$ ETL, (d) forward and reverse scan J-V curves of PSC with C-SnO$_2$ ETL, (e) the statistical distribution diagram of the PCE, and (f) stability of the PSCs stored in an environment of 40 ± 5% humidity (bottom) and in N$_2$ glove box (top).

The cross-sectional SEM image of PSC is shown in Figure 4a. The external quantum efficiency (EQE) of the PSCs with T-SnO$_2$ and C-SnO$_2$ ETLs was tested, as shown in Figure 4b. The results reveal that the PSC with C-SnO$_2$ ETL shows a good optical response with a higher EQE and an integrated current density of 21.78 mA/cm$^2$, while the PSC with T-SnO$_2$ ETL exhibits an integrated current density of 21.30 mA/cm$^2$. Figure 4c,d exhibit the *J–V* characteristics curves of the PSCs with T-SnO$_2$ and C-SnO$_2$ ETLs, and the performance parameters of PSCs for forward and reverse scanning are detailed in inset tables. The PSC optimized with C-SnO$_2$ ETL shows the maximum PCE of 20.34% with a $J_{sc}$ of 23.89 mA/cm$^2$, a $V_{oc}$ of 1.10 V, and an FF of 77.3% for reverse scan, which is higher than that of the T-SnO$_2$-based PSC with a PCE of 19.64% ($J_{sc}$ = 23.67 mA/cm$^2$, $V_{oc}$ = 1.08 V, and FF = 76.5%). Further, PSC with C-SnO$_2$ ETL exhibits reduced hysteresis error. The enhancement in the performance of the PSC was mainly due to the increase in the $V_{oc}$ and FF, which is caused by reduction of nonradiative recombination, improvements in photon capture, and carrier transfer and collection[40]. The statistical distribution diagram of the PCE of the PSCs is shown in Figure 4e and the *J–V* data of 30 devices is shown in Table S1, and the PCE results exhibit that the PSCs with C-SnO$_2$

ETL show good reproducibility and better photovoltaic performance. Time-stability plots of the PSCs are depicted in Figure 4f. The PCE of the PSCs with T-SnO$_2$ and C-SnO$_2$ ETLs, which were stored in an environment of 40 ± 5% relative humidity, decreased to 58% and 61% of its initial value, respectively, after 480 h. There is no significant effect of ETLs observed. The PCE of the device with T-SnO$_2$ ETL, stored in the N$_2$ glove box, decreased to 69% of its initial value after 480 h, while the PCE of the device with C-SnO$_2$ ETL retains over 81% of its initial value after 480 h. This implies that the C-SnO$_2$ layer enhances the stability of the ETL/CsFAMA interface in the PSC.

**CONCLUSION**

In this study, we have introduced the synthesis of (111) facet-engineered C-SnO$_2$ layer as an alternative ETL for PSCs. The C-SnO$_2$ layer improves the surface contact area with the CsFAMA perovskite layer, thereby enhancing the charge transfer dynamics at the interface. Further, the well-matched optical band gap between CsFAMA layer and C-SnO$_2$ ETL enhances the charge extraction efficiency at the interface. In addition, the C-SnO$_2$ promotes the crystal growth in CsFAMA layer with larger nanoparticles and reduces the defect state. As a result of the above optimization, we realize high-performance PSC with PCE of 20.34% for C-SnO$_2$ ETL. The C-SnO$_2$-based devices showed better durability, which exhibited significantly superior performance compared to the PSCs with T-SnO$_2$ ETL. We believe that the proposed C-SnO$_2$ ETL is a promising material for the development of efficient and stable PSCs and offers opportunities for further improvements.

**EXPERIMENTAL METHODS**

**SnO$_2$ precursor solution**

The T-SnO$_2$ precursor solution was prepared by dissolving 31.5 mg of SnCl$_4$.5H$_2$O in 1 mL absolute ethanol. 20.5 μL of acetylacetone (acacH) was added as chelating at room temperature to the solution to yield [acacH]/[Sn] >2. In the presence of an excess of acacH, the hydrolytic stability of the tin-acacchelate complex increases, preventing the progress of further condensation reaction. The nominal hydrolysis ratio of alcoholic solution prepared from SnCl$_4$.5H$_2$O ($h$ = [H$_2$O]/[Sn] adjusted by dropwise addition of 162 μL of DI water to yield $h$ = [H$_2$O]/[Sn] = 105. The solution was stirred at 70 °C for 2 h, giving rise to transparent and stable colloidal solution[42]. The colloidal precursor was deposited at 3000 rpm for 30 s onto the FTO-coated glass. Finally, the deposited film was optimized by an annealing temperature at 180 °C for 1 h in ambient conditions.

The C-SnO$_2$ precursor was prepared by dissolving 0.20 M of SnCl$_4$ in DI water solvent to prepare SnO$_2$ precursor solution. This solution was stirred in an airtight vial at 100 °C for three hours, followed by 3 h stirring at room temperature. A transparent, viscous, and coagulated precursor suspension of tin hydroxide was formed. A 0.22 μm PTFE filter was used to filter the transparent precursor for coating at 4500 rpm for 35 s onto the FTO-coated glass. Finally, the deposited film was optimized by annealing temperature at 145 °C for 1 h in ambient conditions. After cooling down, it was UV ozone treated for 30 min before being used.

**Device fabrication**

The PSCs were fabricated with planar *n–i–p* configuration using fluorine-doped tin oxide coated glass (FTO, 7 Ω/sq)/SnO$_2$/CsFAMA/Spiro-OMeTAD/Au. The FTO-coated glass substrates were cleaned with Hellmanex III (1% in hot deionized water), deionized water, isopropanol for 10 min each, and then subject to UV-O$_3$ treatment for 30 min. For fabricating a C-SnO$_2$ layer, 65 μl of solution was spin coated at 4500 rpm for 30 sec onto the FTO. While the T-SnO$_2$ precursor solution was spin coated at 3000 rpm. The samples were then annealed at 150 °C (for C-SnO$_2$) and 180 °C (for T-SnO$_2$) on a hot plate for 1 h each in ambient conditions to crystallize the SnO$_2$ films and remove organic compounds. For surface treatment of SnO$_2$ layers, a 25 mM SnCl$_2$.2H$_2$O solution in ethanol was spin coated on the FTO/SnO$_2$ samples. For the fabrication of CsFAMA perovskite layer, the precursor solution was prepared by dissolving 507.1 mg of PbI$_2$ and 73.4 mg of PbBr$_2$ in 1 mL of anhydrous DMF and DMSO in a 9:1 (v/v) ratio, stirring at 100 °C for 90 minutes. After cooling, 172 mg of FAI, 22.4 mg of MABr, and 53 μL of a CsI solution (1.5 M in DMSO) were added to the inorganic solution to create the perovskite precursor solution[13]. The precursor solution was then stirred at room temperature for 6 h. The perovskite precursor solution was applied on FTO/SnO$_2$ layer through a two-step spin-coating process. In the first step, the coating was performed at 1000 RPM for 10 secs with a ramping rate of 500 rpm/sec. Subsequently, in the second step, the spin-coating process was carried out at 5000 rpm for 20 secs with a ramping rate of 2000 rpm/sec. During the second step, 200 μL of CB was poured onto the spinning substrate 5 secs prior to the end of spinning program, followed by the annealing at 110 ˚C for 1 h on the hot plate. Spiro-OMeTAD precursor was prepared by dissolving 72.3 mg of spiro-OMeTAD, 17.5 μL of Li-TFSI solution (520 mg in 1 mL ACN), and 28.8 μL of t-BP in 1 mL of CB[13]. 28 μL of spiro OMeTAD solution was then spin-coated upon the perovskite layer at 500 rpm for 3 secs, and 4000 rpm for 30 secs. Finally, an Au electrode was thermally evaporated to a thickness of 100 nm. The deposition rate was set at 1 Å/sec under high vacuum (6.5 × 10$^{-6}$ mbar). The active area of the devices, amounting to 0.09 cm$^2$, was defined using an evaporation mask. To fabricate electron-only devices, PCBM/BCP ETL was used. The precursor solution of PCBM was prepared by dissolving 10 mg in 1 mL of CB and coated at 2000 rpm for 30 s on the perovskite layer. BCP precursor was prepared by dissolving 1 mg in 1 mL of IPA and depositing at 2000 rpm for 30 s on the PCBM layer.

**Material characterization**

The particle size distribution of C-SnO$_2$ precursor was measured using Zeta PALS. The oxidation state of Sn in the C-SnO$_2$ precursor was analysed by Mössbauer spectroscopy in the transmission mode. ATR-FTIR analysis of C-SnO$_2$ precursor was used to identify molecular compounds. HR-TEM analysis was performed to visualize the microscopic structures of SnO$_2$ NPs. XRD and GIXRD measurements were conducted using a Rigaku SmartLab diffractometer that was equipped with a copper Kα anode. The diffractometer operated at a tube output voltage of 45 kV and a current of 30 mA. Chemistry of SnO$_2$ films was examined by Thermo Scientific XPS/UPS system using an Al Kα (λ = 0.83 nm, hυ = 1486.7 eV). X-ray source was operated at 23.5 W, and the data was analyzed using CasaXPS software. The work functions of SnO$_2$ and CsFAMA films were calculated by Thermo-Scientific XPS/UPS system in UPS mode. UV photons are produced using a gas discharge lamp, typically filled with helium. The morphologies of SnO$_2$, CsFAMA films, and cross-sectional image of device were analysed using FE-SEM (ZEISS Ultra55, Mono Carl Zeiss). The absorbance and transmittance spectra of thin films were collected by PerkinElmer LAMBDA 1050+ UV-Vis-NIR spectrometers. The water contact angles on the surface of SnO$_2$ films were measured using Photron FASTCAM Mini UX100. PL spectra of CsFAMA with ETLs were recorded by Confocal Photoluminescence Raman Spectro Microscope (WITec Alpha 300R) at the wavelength of 532 nm. TRPL spectra were recorded at the 405 nm using PicoQuant MicroTime200 time-resolved fluorescence microscope. The EQE measurements of the PSCs were performed by the Oriel IQE-200 series systems. The *J–V* curves were measured using a xenon lamp solar simulator (Newport) with a Keithley 2450 source meter. The measurements were taken under standard test conditions, which include an air mass of 1.5 global (AM 1.5G) spectrum, an irradiance of 100 mW/cm$^2$, and calibration with a standard silicon reference cell.

**ASSOCIATED CONTENT**
**Data Availability Statement**
Data will be made available on request.

**Supporting Information**
The supporting information is available free of charge.

$I$–$V$ curves of $SnO_2$ films, AFM images of $SnO_2$ films, UV-vis absorption and UPS spectra, PL and TRPL spectra of CsFAMA, $J$–$V$ data of 30 independent devices.

**AUTHOR INFORMATION**

**Corresponding Author**
**Ramesh Karuppannan** – Department of Physics, Indian Institute of Science, Bangalore 560012, Karnataka, India
https://orcid.org/0000-0002-8304-6500
Email: kramesh@iisc.ac.in

**Authors**
**Keshav Kumar Sharma** – Department of Physics, Indian Institute of Science, Bangalore 560012, Karnataka, India
Email – keshavsharma@iisc.ac.in
https://orcid.org/0000-0002-6753-2269
**Rohit** – Department of Physics, Indian Institute of Science, Bangalore 560012, Karnataka, India
Email – rohit2023@iisc.ac.in
**Sochannao Machinao** – Department of Physics, Indian Institute of Science, Bangalore 560012, Karnataka, India
Email – sochannaom@iisc.ac.in

**Author Contributions**
K.K. Sharma: writing-original draft, conceptualization, investigation, methodology, data curation and formal analysis. Rohit: methodology and visualization. S. Machinao: data validation and visualization. K. Ramesh: supervision, conceptualization, resources, writing-review and editing and funding acquisition.

**Notes**
The authors declare no competing financial interest.

**ACKNOWLEDGMENT**
We acknowledge support from CeNSE facilities funded by MHRD, MeitY and DST Nano Mission. We thank Indian Science Technology and Engineering facilities Map (I-STEM), a Program supported by Office of the Principal Scientific Adviser to the Govt. of India, for enabling access to the X-ray diffractometer (SmartLab-Rigaku) and Time Resolved Fluorescence Microscope (PicoQuant-MicroTime 200) at Department of Physics, Indian Institute of Science, Bangalore to carry out this work. We would like to thank Dr. Raghavendra Reddy and the UGC-DAE-CSR, Indore, for providing access to the Mössbauer spectrometer facility.

# Supporting Information
## (111) Facet-engineered SnO$_2$ as Electron Transport Layer for Efficient and Stable Perovskite Solar Cells

*Keshav Kumar Sharma, Rohit, Sochannao Machinao, and Ramesh Karuppannan*

Department of physics, Indian Institute of Science, Bangalore 560012, Karnataka, India

Corresponding Author: kramesh@iisc.ac.in

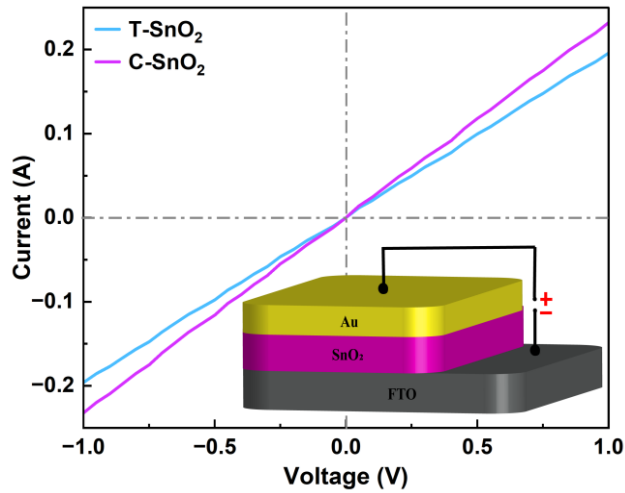

**Figure S1** I-V curves of the T-SnO$_2$ and C-SnO$_2$ thin films.

The optical band gap of the the T-SnO$_2$, C-SnO$_2$ and CsFAMA films are calculated using the UV-vis spectra fit to tauc equation[1]

$$(\alpha h\nu)^n = A(h\nu - E_g)$$

Where $\alpha$ is the absorption coefficient calculated as $\alpha = 2.303\frac{A}{d} = -2.303\frac{1}{d}log\frac{1}{T}$, A is the absorbance, T is the transmittance and d is the thickness of the film. The optical band gap of the T-SnO$_2$, C-SnO$_2$ and CsFAMA films are calculated to 4.01, 4.10 and 1.61 eV, respectively (Figure 2d & S2).

The valance band maximum (E$_v$) was determined by the following equation[2]

$$E_v = (E_{cutoff} - HOS) - h\nu$$

The $E_{cutoff}$ and HOS (higheat occupied state) positions are indicated in Figure 2e & S2. As expected, the $E_v$ values of the T-SnO$_2$, C-SnO$_2$ and CsFAMA films are calculated to be -8.10, -8.04 and -5.40 eV, respectively.

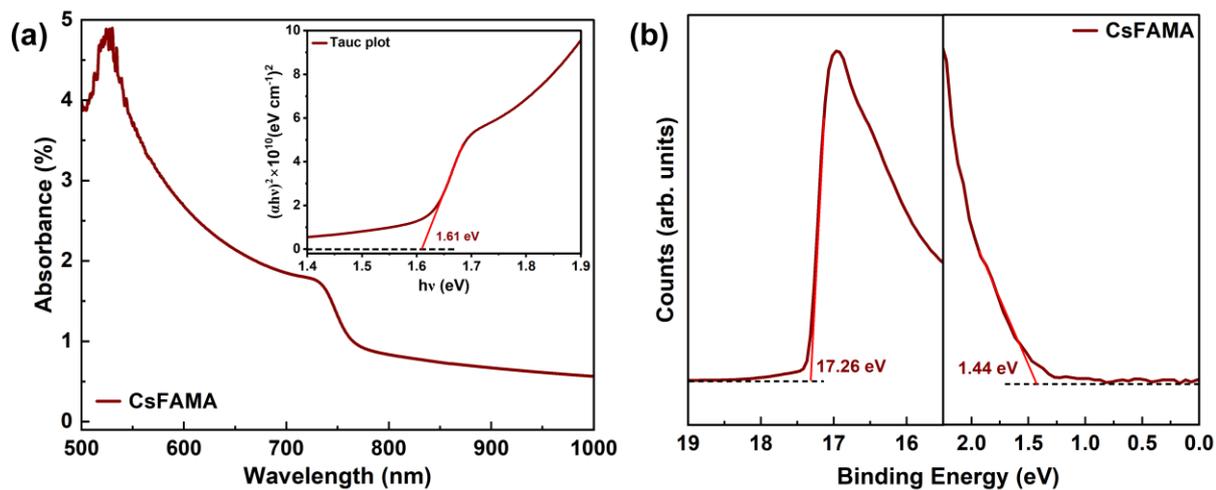

**Figure S2** UV-vis absorption spectrum and UPS spectrum of CsFAMA film.

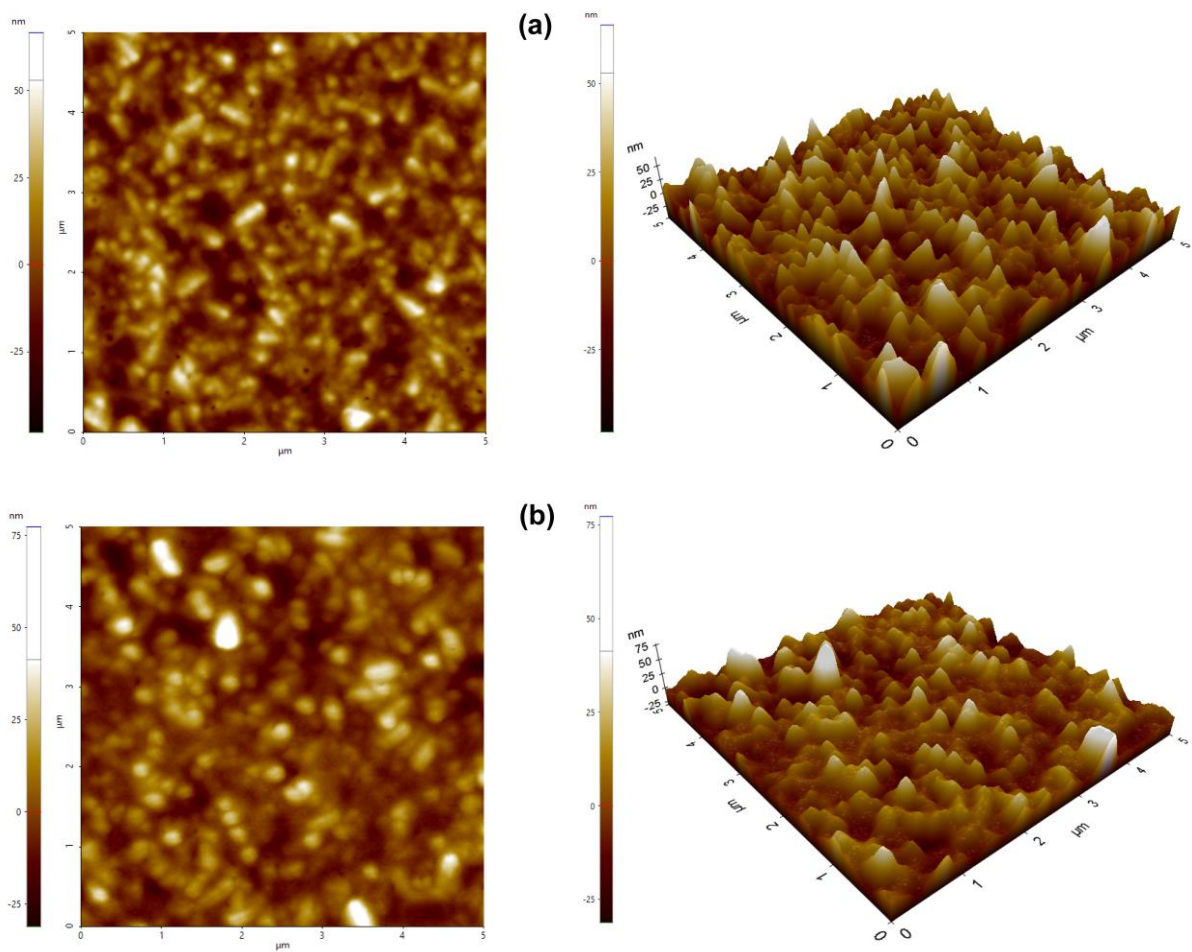

**Figure S3** The 2d and 3d AFM images of (a) T-SnO$_2$ and (b) C-SnO$_2$ thin film on FTO coated glass substrate.

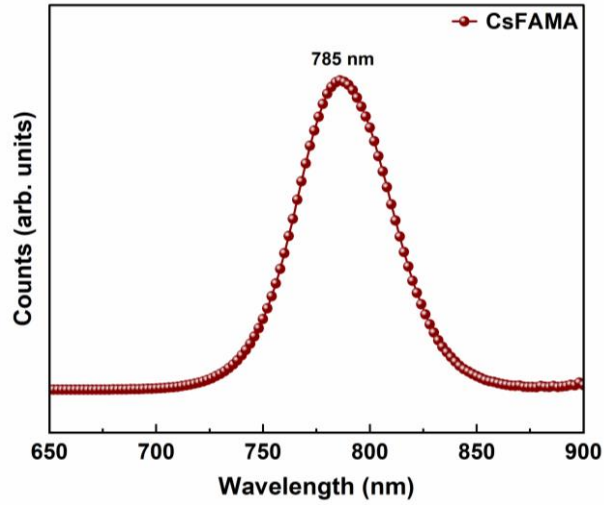

**Figure S4** PL spectrum of pristine CsFAMA film.

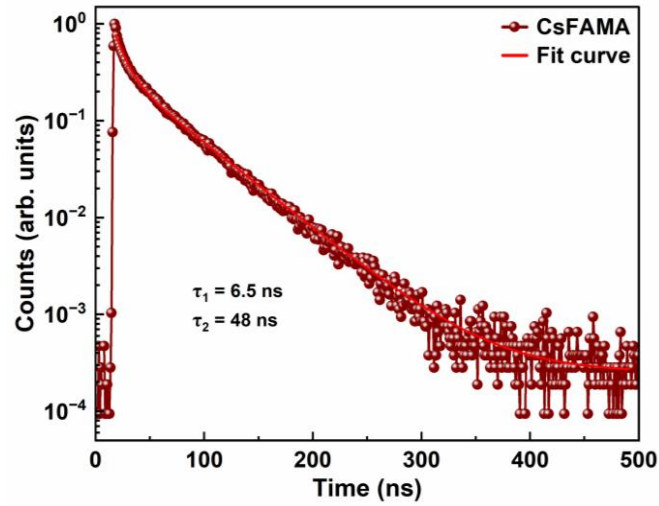

**Figure S5** TRPL spectrum of pristine CsFAMA film.

**Table S1:** The parameters of the SCLC for different electron-only devices.

| Sample | $V_{TFL}$ (V) | $N_{trap}$ (cm$^{-3}$) |
|---|---|---|
| **T-SnO$_2$** | 0.564 | $1.80 \times 10^{16}$ |
| **C-SnO$_2$** | 0.445 | $1.42 \times 10^{16}$ |

The density of trap states was calculated using the following equation[3]

$$N_{trap} = \frac{2\varepsilon_r \varepsilon_0 V_{TFL}}{eL^2}$$

where $V_{TFL}$ is the trap filling limit voltage, L is the thickness of the perovskite layer, $\varepsilon_r$ is the relative dielectric constant of the perovskite layer ($\varepsilon_r \approx 26$), $\varepsilon_0$ is the vacuum dielectric constant, and $e$ is the amount of electric charge.

**Table S2:** The fitting parameters of TRPL spectra

| Sample | $A_1$ | $\tau_1$ (ns) | $A_2$ | $\tau_2$ (ns) | $A_3$ | $\tau_3$ (ns) | $\tau_{ave}$ (ns) |
|---|---|---|---|---|---|---|---|
| CsFAMA | 3.27 | 6.54 | 0.46 | 48.84 | - | - | 28.21 |
| T-SnO$_2$/CsFAMA | 1.26 | 0.75 | 0.70 | 4.06 | 0.10 | 16.75 | 7.38 |
| C-SnO$_2$/CsFAMA | 0.79 | 0.66 | 0.04 | 3.79 | 0.005 | 17.45 | 3.21 |

The average lifetime was determined by the following equation[4]

$$<\tau_{ave}> = \frac{\sum_{i=1,2,3} A_i \tau_i^2}{\sum_{i=1,2,3} A_i \tau_i}$$

where $A_i$ is the pre-exponential factor and $\tau_i$ is the average lifetime.

**Table S3:** *J–V* data of 30 independent devices.

| T-SnO$_2$ | $V_{OC}$ (V) | $J_{SC}$ (mA.cm$^{-2}$) | FF (%) | PCE (%) | C-SnO$_2$ | $V_{OC}$ (V) | $J_{SC}$ (mA.cm$^{-2}$) | FF (%) | PCE (%) |
|---|---|---|---|---|---|---|---|---|---|
| 1 | 1.08 | 23.67 | 76.5 | 19.64 | 1 | 1.10 | 23.89 | 77.3 | 20.34 |
| 2 | 1.08 | 23.53 | 76.3 | 19.38 | 2 | 1.10 | 23.76 | 76.9 | 20.09 |
| 3 | 1.07 | 23.87 | 75.8 | 19.36 | 3 | 1.10 | 23.89 | 75.3 | 19.78 |
| 4 | 1.09 | 23.69 | 74.1 | 19.13 | 4 | 1.11 | 23.54 | 75.6 | 19.75 |
| 5 | 1.08 | 23.45 | 74.3 | 18.81 | 5 | 1.10 | 23.48 | 76.3 | 19.70 |
| 6 | 1.07 | 23.58 | 73.8 | 18.62 | 6 | 1.10 | 23.64 | 75.1 | 19.53 |
| 7 | 1.08 | 23.33 | 72.9 | 18.55 | 7 | 1.10 | 23.67 | 74.9 | 19.50 |
| 8 | 1.07 | 23.68 | 73.2 | 18.54 | 8 | 1.09 | 23.93 | 74.7 | 19.48 |
| 9 | 1.08 | 23.33 | 73.1 | 18.42 | 9 | 1.10 | 23.83 | 74.3 | 19.47 |
| 10 | 1.06 | 23.59 | 73.3 | 18.33 | 10 | 1.11 | 23.06 | 75.5 | 19.32 |
| 11 | 1.07 | 23.47 | 72.6 | 18.24 | 11 | 1.11 | 23.23 | 74.8 | 19.29 |
| 12 | 1.06 | 23.77 | 72.3 | 18.22 | 12 | 1.10 | 23.63 | 74.2 | 19.28 |
| 13 | 1.06 | 23.46 | 72.5 | 18.03 | 13 | 1.11 | 23.14 | 74.8 | 19.21 |
| 14 | 1.07 | 23.11 | 72.1 | 17.83 | 14 | 1.09 | 23.67 | 74.0 | 19.06 |
| 15 | 1.06 | 23.34 | 71.6 | 17.72 | 15 | 1.10 | 23.46 | 73.6 | 18.99 |
| 16 | 1.05 | 23.53 | 71.5 | 17.66 | 16 | 1.09 | 23.46 | 73.3 | 18.74 |
| 17 | 1.07 | 23.48 | 70.3 | 17.67 | 17 | 1.08 | 23.54 | 73.5 | 18.68 |
| 18 | 1.06 | 23.41 | 70.7 | 17.55 | 18 | 1.10 | 23.14 | 73.1 | 18.61 |
| 19 | 1.06 | 23.64 | 69.5 | 17.42 | 19 | 1.11 | 23.06 | 72.6 | 18.58 |
| 20 | 1.08 | 23.45 | 68.5 | 17.35 | 20 | 1.09 | 23.34 | 72.9 | 18.54 |
| 21 | 1.07 | 23.61 | 68.4 | 17.28 | 21 | 1.09 | 23.27 | 72.8 | 18.46 |
| 22 | 1.07 | 23.27 | 69.1 | 17.21 | 22 | 1.08 | 23.45 | 72.6 | 18.38 |
| 23 | 1.06 | 23.64 | 66.9 | 16.76 | 23 | 1.11 | 22.86 | 72.3 | 18.34 |
| 24 | 1.05 | 24.34 | 64.7 | 16.54 | 24 | 1.10 | 23.56 | 70.3 | 18.21 |
| 25 | 1.06 | 23.71 | 67.1 | 16.87 | 25 | 1.09 | 23.85 | 69.1 | 17.96 |
| 26 | 1.06 | 23.49 | 66.1 | 16.46 | 26 | 1.10 | 23.66 | 68.3 | 17.76 |
| 27 | 1.06 | 23.12 | 66.3 | 16.24 | 27 | 1.09 | 24.14 | 66.2 | 17.42 |
| 28 | 1.08 | 22.76 | 64.2 | 15.78 | 28 | 1.10 | 23.87 | 65.4 | 17.18 |
| 29 | 1.05 | 23.22 | 64.6 | 15.75 | 29 | 1.10 | 23.36 | 65.8 | 16.90 |
| 30 | 1.05 | 23.05 | 63.6 | 15.39 | 30 | 1.08 | 23.06 | 65.3 | 16.26 |